\documentclass{article}
\usepackage{geometry}
\usepackage{graphicx}
\usepackage{cite}
\usepackage{amsmath}

\usepackage{subfigure}

\usepackage{tkz-linknodes}

\usepackage{soul}
\renewcommand{\ul}{}

\begin{document}
\title{Arbitrary coupling ratio multimode interference couplers in Silicon-on-Insulator}

\author{Jos\'e~David~Dom\'enech$^1$, Javier~S.~Fandi\~no$^2$, Bernardo~Gargallo$^2$ and Pascual~Mu\~noz$^{1,2}$ \\ 
\\
\small $^1$VLC Photonics S.L., C/ Camino de Vera s/n, \\ 
\small Valencia 46022, Spain e-mail: david.domenech@vlcphotonics.com \\
\small $^2$Optical and Quantum Communications Group, iTEAM Research Institute, \\ 
\small Universitat Polit\`ecnica de Val\`encia, C/ Camino de Vera s/n, \\ 
\small Valencia 46022, Spain e-mail: pmunoz@iteam.upv.es.
}


\maketitle

\begin{abstract}
In this paper we present the design, manufacturing, characterization and analysis of the coupling ratio spectral response for Multimode Interference (MMI) couplers in Silicon-on-Insulator (SOI) technology. The couplers were designed using a Si rib waveguide with SiO$_2$ cladding, on a regular 220~nm film and 2 $\mu$m buried oxide SOI wafer. A set of eight different designs, three canonical and five using a widened/narrowed coupler body, have been subject of study, with coupling ratios 50:50, 85:15 and 72:28 for the former, and 95:05, 85:15, 75:25, 65:35 and 55:45 for the latter. Two wafers of devices were fabricated, using two different etch depths for the rib waveguides. A set of six dies, three per wafer, whose line metrology matched the design, were retained for characterization. The coupling ratios obtained in the experimental results match, with little deviations, the design targets for a wavelength range between 1525 and 1575~nm, as inferred from spectral measurements and statistical analyses. Excess loss for all the devices is conservatively estimated to be less than approximately 2~dB. All the design parameters, body width and length, input/output positions and widths, and tapers dimensions are disclosed for reference.
\end{abstract}


\section{Introduction}
Optical couplers are perhaps one of the most basic and most used among the building blocks for photonic integrated circuits (PICs) in all currently available technology platforms \cite{munoz_icton2013}. Different integrated implementations exist (see \cite{hunsperger}), and they are usually compared according to their coupling constant and operational wavelength range. Among all of them, the Multimode Interference (MMI) coupler is mostly used in high index contrast PIC technologies, such as III-V and group IV materials, since it is in general more compact and preserves the coupling constant over a wide wavelength range. Since its inception by Ulrich in 1975 \cite{ulrich1975}, and the demonstrations of MMIs as we know them today carried out by Pennings and co-workers \cite{pennings1991}, multitude of papers have studied the different aspects of these very versatile devices: fundamental theory and design rules for the so called canonical MMIs, by Soldano \cite{soldano} and Bachmann \cite{bachmann94,bachmann95}; design rules and experimental demonstrations of widened/narrowed body MMIs for arbitrary coupling ratios at a single wavelength by Besse \cite{besse96}, with reconfiguration using thermal tuning by Leuthold \cite{leuthold01}; tolerance analysis by Besse \cite{besse94}; design optimizations for different technologies by Halir \cite{halir}; library of experimentally demonstrated 50:50 Silicon-on-Insulator couplers \cite{zhou}, to name a few. \ul{There are other means of implementing couplers with arbitrary ratio that make use of additional structures, as for instance the combination of two MMIs in a Mach-Zehnder Interferometer (MZI) like structure recently proposed by Cherchi} \cite{art:cherchi14}.

In this paper we report on the design and experimental demonstration of arbitrary coupling ratio MMIs following the design rules by Besse and co-workers \cite{besse96}, supported by Beam Propagation Method (BPM) commercial software optimizations \cite{phoenix}, on a Silicon-on-Insulator (SOI) platform. \ul{Complimentary to previous works available in the literature this paper presents:, a) all} the design parameters required to obtain broadband (1525-1575~nm) coupling ratio, with modest excess loss, for canonical 50:50, 85:15 and 72:28 MMIs, as well as for widened/narrowed 95:05, 85:15, 75:25, 65:35 and 55:45 MMIs are disclosed\ul{; b) spectral traces demonstrating the otherwise well known theoretically broadband operation of these devices; c) statistics for the coupling ratio variations in the operational wavelength range, that may be of use to perform variational analysis of more complex on-chip devices, circuits and networks based on these MMIs; d) explanation on how measurement deviations, due to variations in the in/out coupling to/from the chip, can bias the coupling ratio results, and e) measurements to infer the reproducibility, die to die and wafer to wafer, of the responses.} These reference designs, experimentally validated, \ul{together with the statistical variations and reproducibility information,} can be used as starting point for other designers \ul{and researchers of these devices, and of more complex chip networks employing them,} on SOI platforms. 

\section{Design}
The design of all the MMIs was carried out in three steps: i) cross-section analysis and 2D reduction, ii) analytic approach and iii) numerical BPM optimization. The cross section consists of a buried oxide layer of 2 microns height, capped with a 220~nm Si layer and a SiO$_2$ over-cladding. Rib waveguides, with 130~nm etch depth from top of the Si layer, were used in the design stage. For the same lithographic resolution, rib waveguides provide more robust MMIs than strip waveguides, owing to the fact that wider waveguides are required to support the same number of modes \cite{soldano}. This comes at the cost of increased footprint and some additional design refinements are required to minimize the MMI imbalance and excess loss\cite{halir}\cite{hill}, besides the complexity of two mask level fabrication described in \cite{thomson2010low}. The latter trade-off is common in applications where the coupling constant needs to be set very precisely, for instance in very small free spectral range Mach-Zehnder interferometers (MZI), to compensate for the significantly larger loss difference between the long and short interferometer arms \cite{bogaerts2010silicon}. Moreover it is determinant for on-chip reflectors based on Sagnac interferometers, where the reflectivity is solely determined by the coupling ratio of the coupler in the interferometer \cite{munoz2011sagnac}.

Firstly, for the cross-section analysis a film-mode matching mode solver was used \cite{phoenix}. The wavelength dependence of the refractive indices was included in the solver (see the Appendix). For a given MMI width, the first and second mode propagation constants, $\beta_0$ and $\beta_1$ respectively, were found for a wavelength of 1.55 $\mu$m for TE polarization, and the beat length $L_{\pi} = \pi / (\beta_0-\beta_1)$ was computed from these. For the case of all the MMIs subject to design, the body width was set to 10~$\mu$m. The effective indices for the first and second mode given by the solver are n$_{eff,0}$=2.84849 and n$_{eff,1}$=2.84548. Therefore the beat length results into L$_{\pi}$=257.61~$\mu$m. In order to later use a 2D BPM method, the cross-section was reduced vertically to a 1D waveguide using the effective index method (EIM) \cite{buus}. EIM was firstly used to derive the 1D effective index for the core region, and then the effective index left/right to the core was calculated by numerically solving (with a bisection method) for the 1D modes of the reduced structure to match the previously calculated $L_{\pi}$ on the 2D cross-section.

Secondly, analytic design rules for canonical \cite{soldano} and arbitrary coupling ratio \cite{besse96} MMIs were used. These rules provide, for a given MMI width, an analytic approximation for the MMI body length, \ul{named L$^0$}, from the previously calculated $L_{\pi}$, and for the case of arbitrary ratio, the width variation and body geometry (named type A, B, C and D in \cite{besse96}). \ul{For completeness, the analytic approximations for the MMI lengths are reproduced here:}

\begin{eqnarray}
L^0_A &=&  \delta_W^A \frac{1}{2} \left(3 L_\pi \right) \label{eq:La}\\
L^0_{B,Bsym} &=& \delta_W \frac{1}{3} \left(3 L_\pi \right) \label{eq:Lb}\\
L^0_C &=& \delta_W \frac{1}{4} \left(3 L_\pi \right) \label{eq:Lc}\\
L^0_D &=& \delta_W \frac{1}{5} \left(3 L_\pi \right) \label{eq:Ld}
\end{eqnarray}
\ul{Where 3$L_\pi$ is the distance for the first direct (not mirrored) image} \cite{soldano}, \ul{$\delta_W^A = 1-\Delta W/W$ and $\delta_W = 1-2\Delta W/W$. Note in the case of rectangular body, $\Delta W$=0 and the last two expresions are equal to 1.} Up to this stage, only the MMI body width and a first guess for the length are set.

\begin{figure}
  {\par\centering
   \subfigure[]{\resizebox*{0.48\textwidth}{!}{\includegraphics*{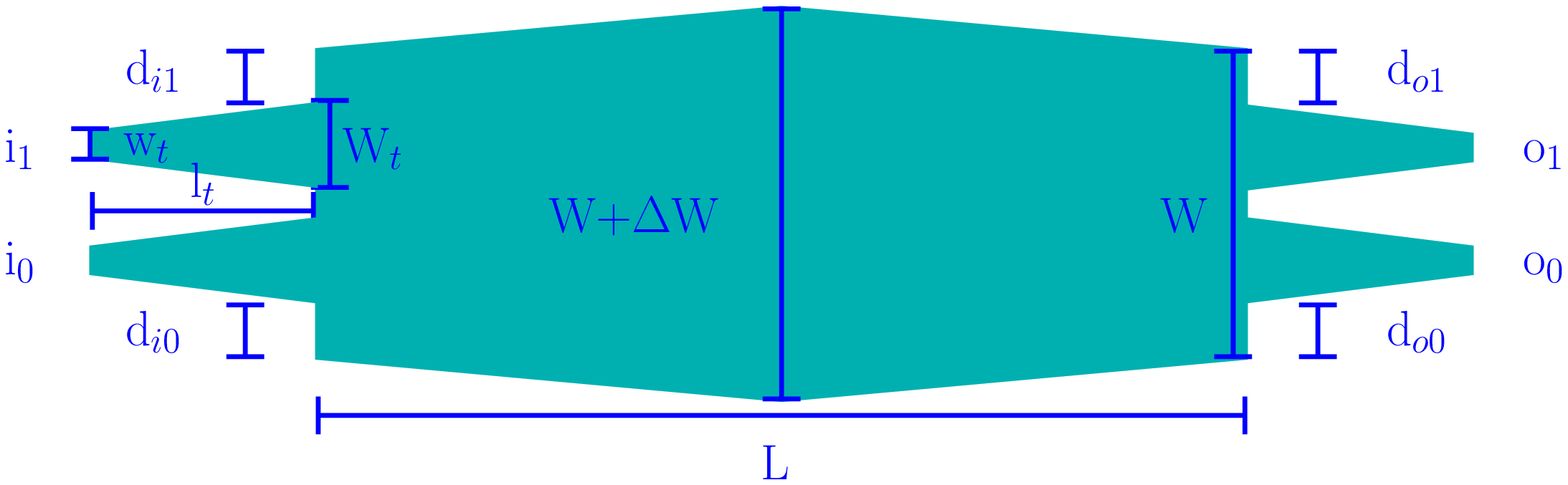}}}
   \subfigure[]{\resizebox*{0.48\textwidth}{!}{\includegraphics*{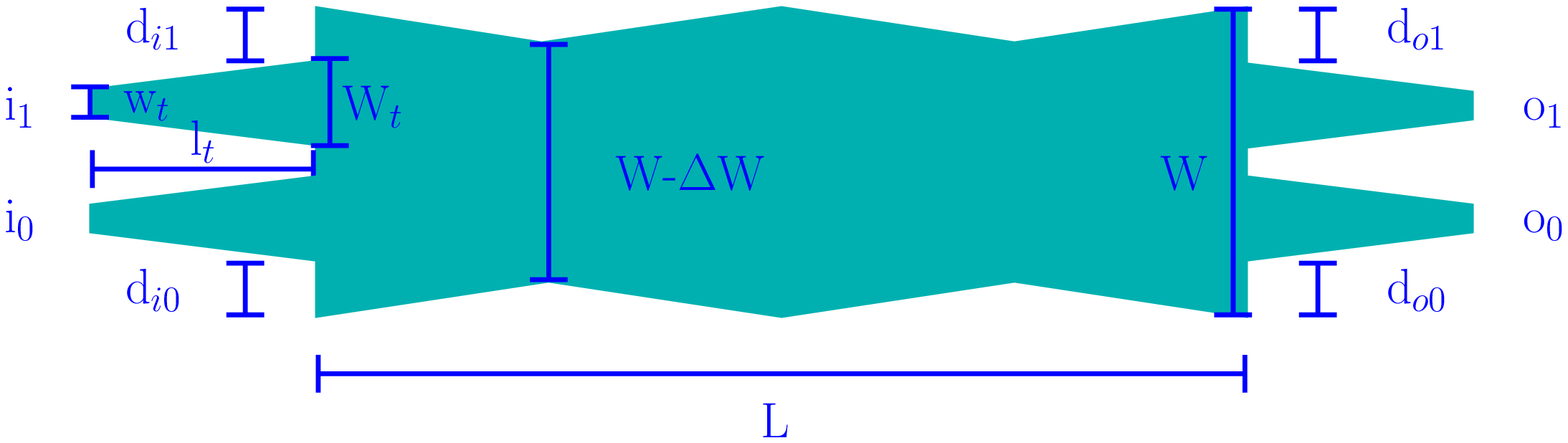}}}
   \caption{Multimode Interference coupler geometries employed in this work, (a) Type A, C, D layouts and (b) Type B Symmetrized layout. Abbreviations: $L$ and $W$, MMI body length and width respectively; $d_{io}$ distance of input/output waveguides from the edges of the MMI body; $l_t$ input/output waveguide taper length; $w_t$ and $W_t$ input/output taper narrow and wide side widths, respectively.}
  \label{fig:mmi_sketch}
  }
\end{figure}

The final step consists of using BPM for a MMI having input/output tapered waveguides. Tapers are required to minimize the MMI excess loss, imbalance and reflections as described in \cite{halir}\cite{hill}. Hence, BPM is used to find iteratively both the MMI length and the input/output tapers width. \ul{The optimization process has as target to minimize the coupler imbalance, i.e. that the ratios at both outputs match the target, and to maximize the overall optical power with respect to the input, i.e. to minimize the excess loss. To do so, the BPM is equiped with mode overlap monitors at the output waveguides. They provide the amplitude and phase for the overlap of the output field with the fundamental mode of the waveguide. The optimization process starts with a fixed set of taper width and MMI length. The starting taper width was set to 3~$\mu$m. The taper length was set to 50.0~$\mu$m, which is sufficiently large for adiabatic linear tapering as per} \cite{art:Fu14}\ul{. The MMI length was set to the values obtained through the aforementioned analytic formulas. They provide an MMI length that does not account the tapering of the input/outputs, which in turn modifies the propagation conditions in the multimode waveguide. Therefore, for the initial guess of taper width, the length of the MMI is solved numerically in a first step. Next, the width of the taper is varied. Both parameters are iteratively changed following update and minimization numerical methods commonly now, until a solution is found for the coupling ratios, having as stop condition a tolerance of 0.01.} The optimization was performed firstly for $\lambda$ = 1.55 $\mu$m, and finally cross-checked for the design wavelength interval, 1.525-1.575 $\mu$m. The body shapes and parameters for the MMIs are given in Fig.~\ref{fig:mmi_sketch} and Table~\ref{tab:mmi_param}. \ul{The parameters subject to numerical optimization are marked in the table with the $^*$ symbol.} 

\ul{Note the optimization process yields shorter MMI body lengths than those provided by the analytic expressions in Eqs.} (\ref{eq:La})-(\ref{eq:Ld}). \ul{This can be explained in terms of the underlying physics as follows. The analytic expressions provide the length for a perfectly rectangular body. Including input/output tapers perturbs the rectangular body shape in the longitudinal dimension, producing multimode propagation in a length larger than the canonical rectangular length. The effect is similar to having a perfectly rectangular body, but with increased length. Hence, to compensate this extra propagation length in the tapers, the body length needs to be reduced.}

    \begin{table}
      \begin{center}
    \begin{tabular}{|c|c|c|c|c|} \hline
      \multicolumn{5}{|c|}{\bfseries MMI Design Parameters} \\ \hline
      Id &      \#1 &      \#2 &      \#3 &      \#4 \\ \hline\hline
      Ratio & 50:50 & 85:15 & 95:05 & 85:15 \\ \hline
      Type & A & C & B Sym & B Sym \\ \hline
      L$^0$ & $\delta_W^A$(1/2)L$_\pi$ & $\delta_W$(3/4)L$_\pi$ & $\delta_W$(3/3)L$_\pi$ & $\delta_W$(3/3)L$_\pi$ \\
            & 128.81 & 193.21 & 220.40  & 192.75 \\ \hline
      L$^*$ & 122.96 & 184.95 & 211.95 & 184.55 \\ \hline
      W & 10.00 & 10.00 & 10.00 & 10.00 \\ \hline
      $\Delta$W &0&0& 0.72 & 1.26 \\ \hline
      d$_{i0}$,d$_{i1}$ & 1.90 & 0.83 & 1.97 & 1.97 \\ \hline
      d$_{o0}$,d$_{o1}$ & 1.90 & 0.83 & 1.97 & 1.97 \\ \hline
      l$_t$ & 50.00 & 50.00 & 50.00 & 50.00 \\ \hline
      w$_t$ & 0.45 & 0.45 & 0.45 & 0.45 \\ \hline
      W$_t^0$ & 3.00 & 3.00 & 3.00 & 3.00 \\ \hline
      W$_t$$^*$ & 2.75 & 3.35 & 2.7 & 2.7 \\ \hline
      Body  & Single  & Single  & Double  & Double  \\ \hline\hline
      Id &      \#5 &      \#6 &      \#7 &      \#8 \\ \hline\hline
      Ratio & 75:25 & 65:35 & 55:45 & 72:28 \\ \hline
      Type & C & D & D & D \\ \hline
      L$^0$ & $\delta_W$(3/4)L$_\pi$  & $\delta_W$(3/5)L$_\pi$  & $\delta_W$(3/5)L$_\pi$  & $\delta_W$(3/5)L$_\pi$ \\
            & 257.49 & 177.66 & 208.55 & 154.57 \\ \hline
      L$^*$ & 247.18 & 170.36 & 200.04 & 147.76 \\ \hline
      W & 10.00 & 10.00 & 10.00 & 10.00 \\ \hline
      $\Delta$W & 3.26 & 1.5 &  3.34 & 0 \\ \hline
      d$_{i0}$,d$_{i1}$ & 1.11 & 0.60,2.60 & 0.61,2.61 & 0.75,2.75 \\ \hline
      d$_{o0}$,d$_{o1}$ & 1.11 & 2.60,0.60 & 2.61,0.61 & 2.75,0.75 \\ \hline
      l$_t$ & 50.00 & 50.00 & 50.00 & 50.00 \\ \hline
      w$_t$ & 0.45 & 0.45 & 0.45 & 0.45 \\ \hline
      W$_t^0$ & 3.00 & 3.00 & 3.00 & 3.00 \\ \hline
      W$_t$$^*$ & 2.83 & 2.83 & 2.80 & 2.50 \\ \hline
      Body & Single  & Single  & Single  & Single  \\ \hline
    \end{tabular}
    \vspace{2mm}
    \caption{\label{tab:mmi_param} MMI design parameters, lengths and widths in $\mu m$. Parameters marked with the symbol $*$ were subject to numerical optimization. $\delta_W = 1-2\Delta W/W$. $\delta_W^A = 1-\Delta W/W$}
    \end{center}
  \end{table}

\section{Experimental results}
\subsection{Fabrication}
\begin{figure}
  {\par\centering
   \resizebox*{0.48\textwidth}{!}{\includegraphics*{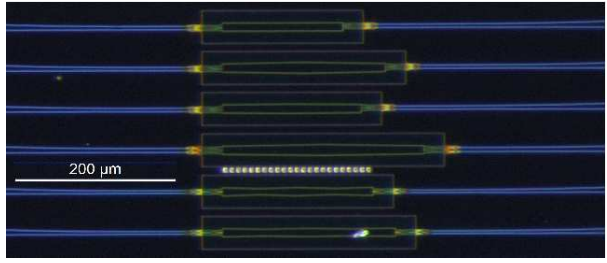}}
   \caption{Chip photograph for the fabricated MMI devices, from bottom to top MMIs \#3-\#8, i.e. 95:05, 85:15, 75:25, 65:35, 55:45 (non canonical) and 72:28 (canonical) splitting ratios.}
  \label{fig:mmi_chip}
  }
\end{figure}

The designs were fabricated on two different wafers, named A and B from now onwards, using a 248~nm CMOS line Multi-Project Wafer at the Institute of Micro-Electronics, Singapore. The difference between wafers A and B was the etch depth for the rib waveguides, 130~nm (as per design) and 160~nm from top of the Si film, respectively. From the dies delivered by the fab, those exhibiting metrology (grating line width, grating space width, waveguide width) close to target were retained for measurements. The target grating line and space were both 315~nm. Metrology shows grating line in [321,333]~nm, and grating space in [296-312]~nm. The target waveguide width for the process calibrations at the fab was 500~nm, and metrology shows widths in [520,541]~nm. The number of dies with metrology in these ranges amounted for 3 dies per wafer, namely A1, A2 and A3 for wafer A, and B1, B2, B3 for wafer B. A picture of the fabricated devices is shown in Fig.~\ref{fig:mmi_chip}. All the layouts included focusing grating couplers (FGC) to couple light vertically into/out from the chips \cite{fgc}. Both the FGCs and the waveguides connected to the MMIs supported only TE polarization.

\subsection{Characterization setup}
The characterization setup consists of a set of three motorized positioners. Two of them are used for holding the fibers vertically at the right angle to couple light into the FGCs (10$^{\circ}$ from the normal to the chip surface), whereas the third one holds the sample on top of a thermally controlled (25~$^{\circ}$C) vacuum chuck. A CCD camera vision system is also mounted in a motorized stage and a LED lamp is used for illumination. For the measurements, the fibers are aligned manually in two steps. Firstly, the fibers are approximated to the FGC locations by visual inspection using the live images from the camera. The approximated height location can be obtained when the fiber and its shadow overlap. Secondly, a broadband light source is connected to one of the fibers, whereas a power meter is connected at the end of the other. Hence, the positions of the input and output fibers are optimized with the motorized stages to obtain the maximum power. A further detailed description of this setup and procedure can be found in \cite{domenech_phd}. After the fibers alignments are optimized, an Optical Spectrum Analyzer (OSA) is used to record the spectra with a resolution of 10~pm.
\begin{figure}
  {\par\centering
   \resizebox*{0.48\textwidth}{!}{\includegraphics*{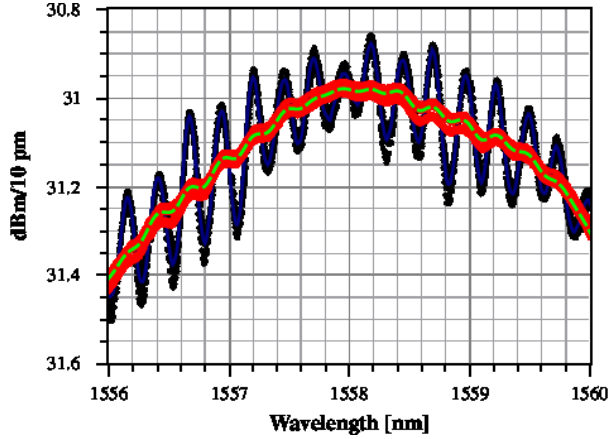}}
   \caption{Spectral traces recorded consecutively for one straight waveguide. The black dots correspond to 8 traces, while their average is shown in marine blue solid line. The red dots correspond to the former 8 traces smoothed with a moving average of 71 points, and the green dashed line to its average.}
  \label{fig:8sws}
  }
\end{figure}

\subsection{Test structures and stability}
Prior to measuring and processing the target devices, straight waveguides were measured in order to gather information on the different features observed in preliminarily recorded traces. Referring to Fig.~\ref{fig:8sws}, a single straight waveguide was measured repeatedly, and a set of 8 consecutive traces was obtained. These are shown in black dots in the figure, with the average in marine blue solid line. Some Fabry-P\'erot (FP) like ripples were observed, with a separation between peaks of 0.26~nm. These are attributed to reflections occurring elsewhere, and that are otherwise not present in the spectrum of the optical source.
Thus a moving average of 71 points (traces were recorded with a spectral resolution of 10~pm, this corresponds to 0.71~nm, approximately twice $\Delta \lambda$) was applied to the 8 traces, in order to clean the FP peaks. The results are shown in Fig.~\ref{fig:8sws} with red dots for the 8 traces, whereas the average for them is shown in a dashed green line. Hence, all spectral traces recorded for the MMIs are to be smoothed as described, before using them in the calculations of coupling ratios described in the next subsection. As a final remark on stability, each trace involved a sweep in the OSA of 20~seconds. From Fig.~\ref{fig:8sws} a good setup stability can be inferred, i.e. very little power variations due to mechanical issues, as the fiber holders, translation stages, tabletop and other during the time it took to acquire these 8 traces (close to 3 minutes) is attained. Though not shown, minor drifts in the alignments started to happen right after the time for the 8 traces. For the MMIs a single trace is collected per output, therefore, it is not subject to the mechanical drifts of the measurement setup.
\begin{figure}
  {\par\centering
  \subfigure[MMI\#2]{\resizebox*{0.45\textwidth}{!}{\includegraphics*{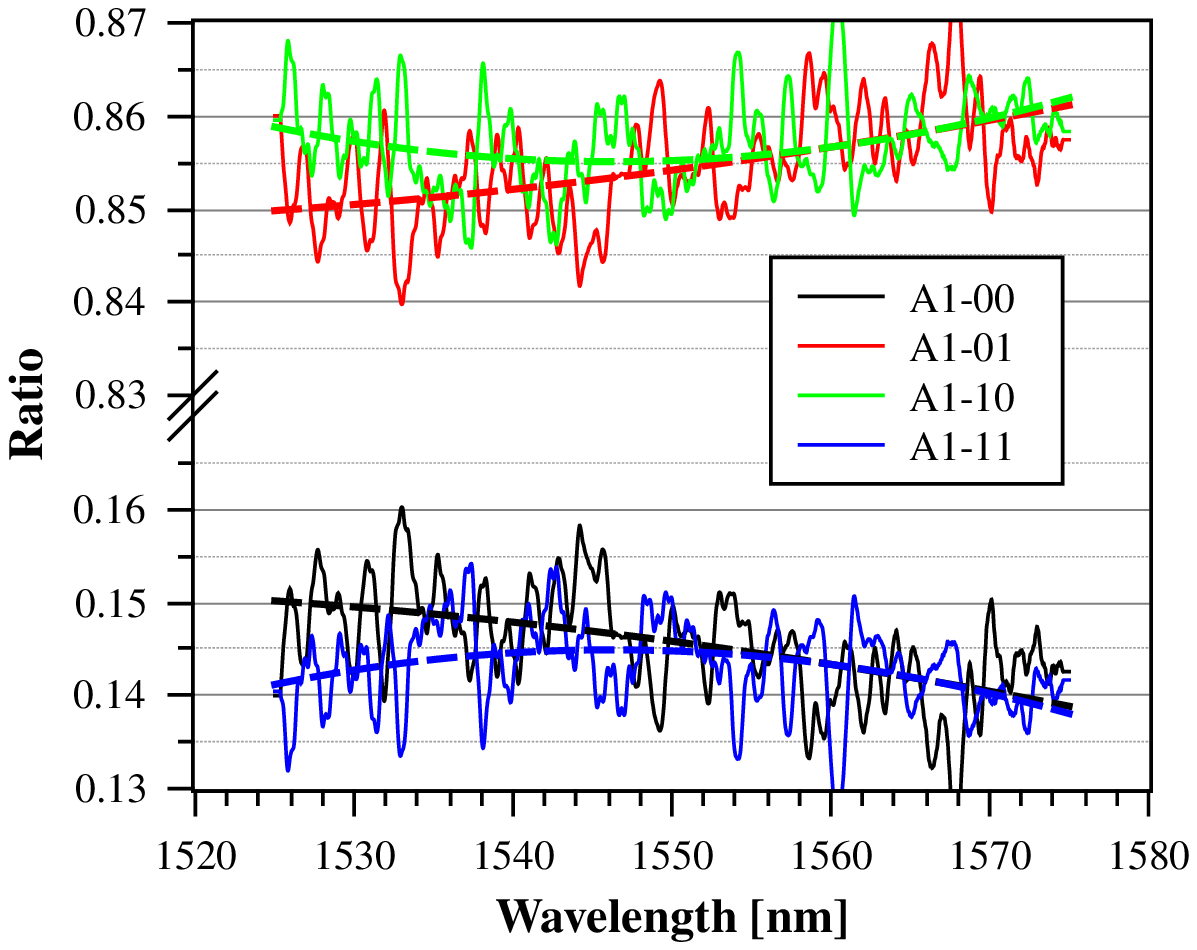}}}
  \subfigure[MMI\#3]{\resizebox*{0.45\textwidth}{!}{\includegraphics*{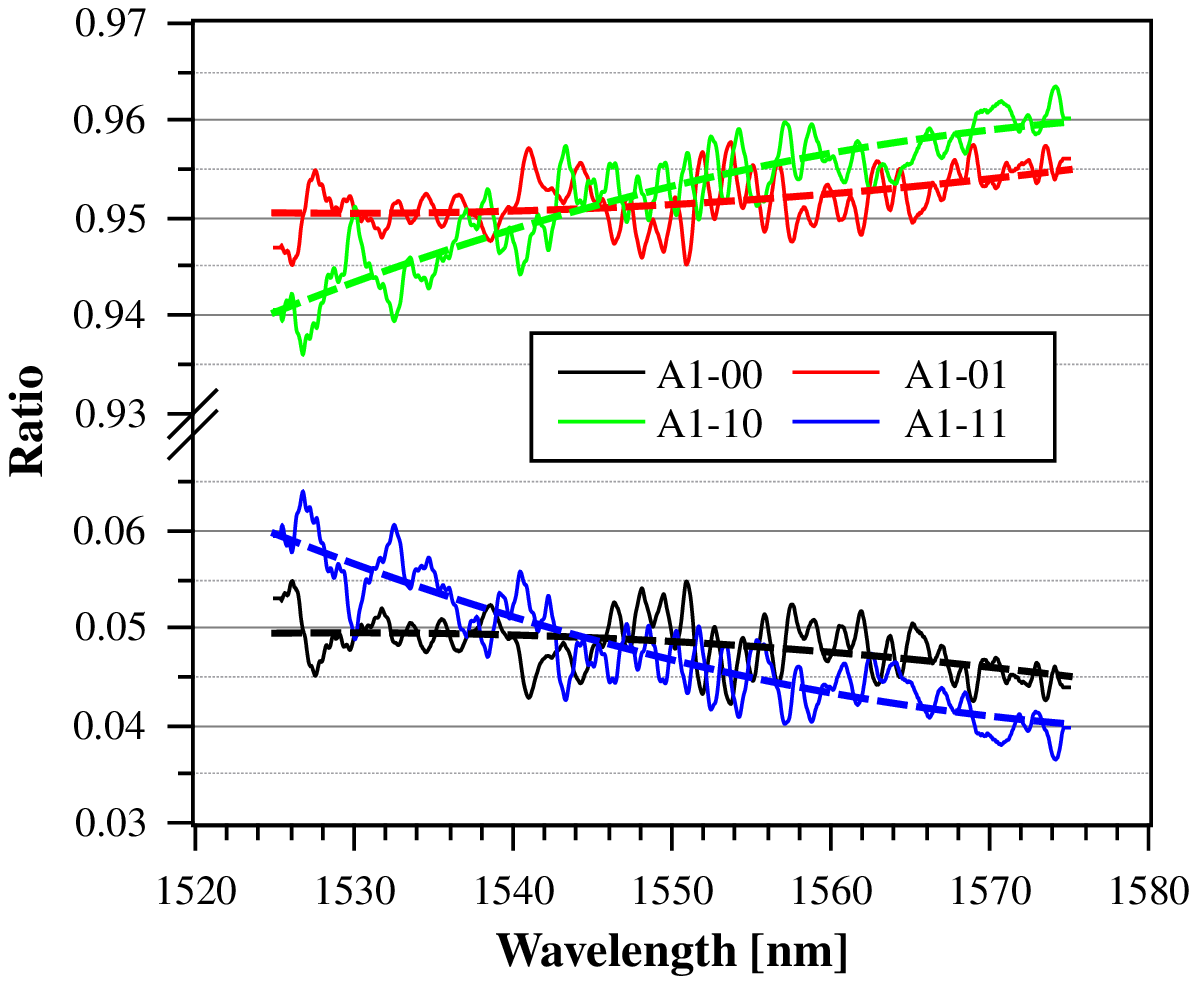}}}
  \caption{Spectral coupling ratio for canonical 85:15 MMI (a) and narrowed 95:05 MMI (b) in solid lines, with second order polynomial fit in dashed thick lines.}
  \label{fig:mmi_spectra}
  }
\end{figure}

\subsection{MMIs Coupling ratio}
\begin{figure}
{\par\centering


    \subfigure[MMI\#1]{\resizebox*{0.35\textwidth}{!}{\includegraphics*{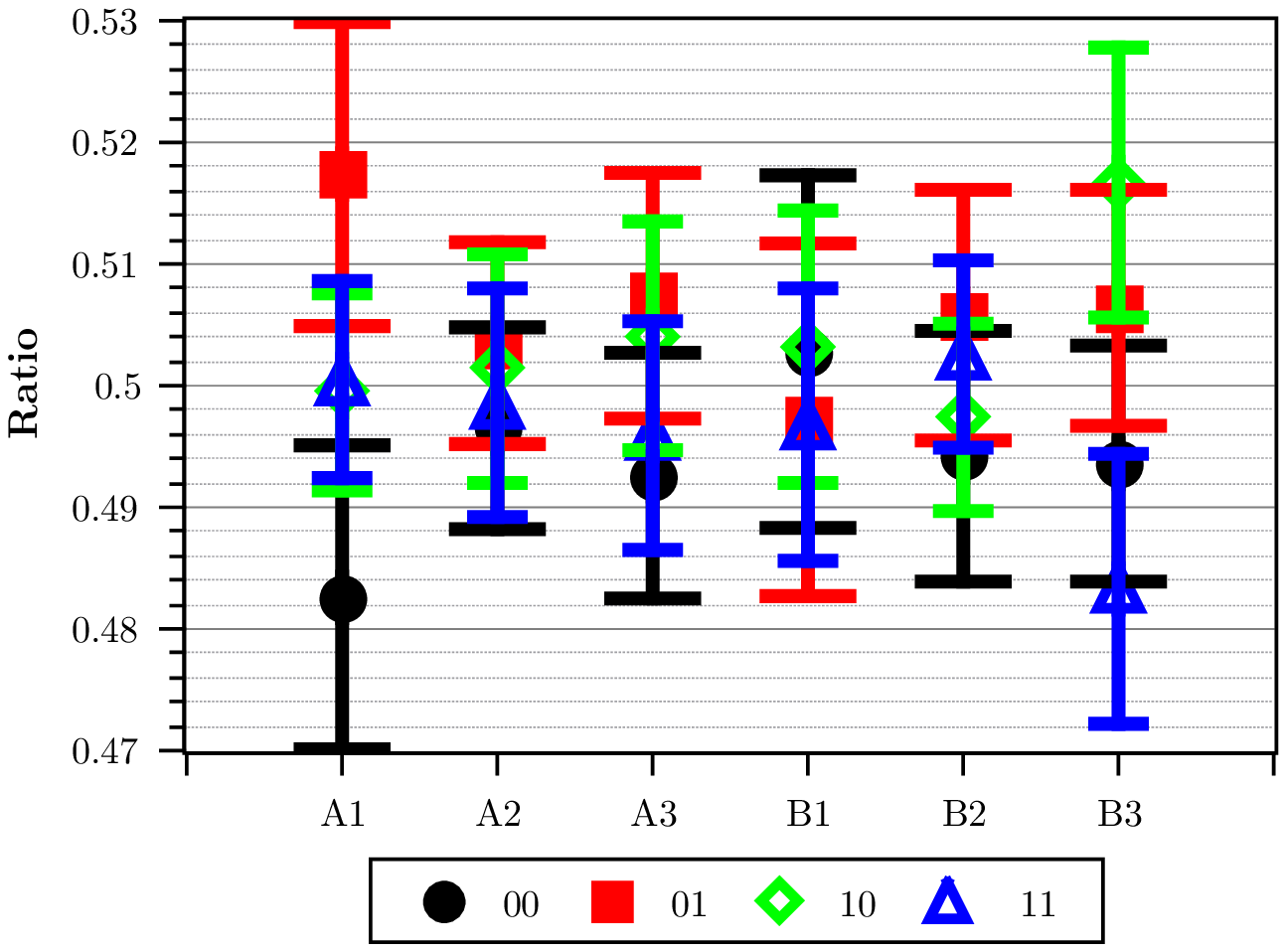}}}
    \subfigure[MMI\#2]{\resizebox*{0.35\textwidth}{!}{\includegraphics*{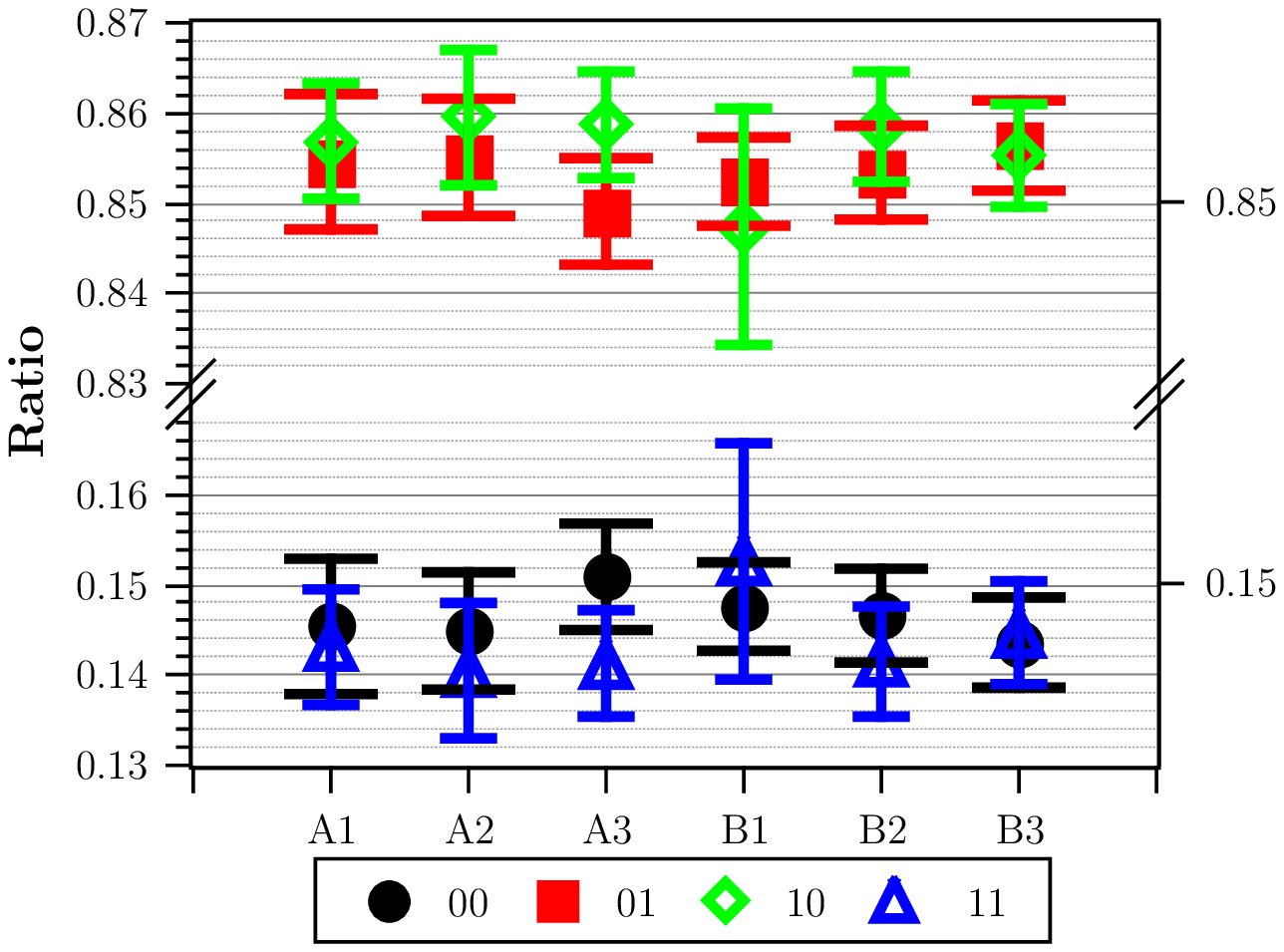}}}

    \subfigure[MMI\#3]{\resizebox*{0.35\textwidth}{!}{\includegraphics*{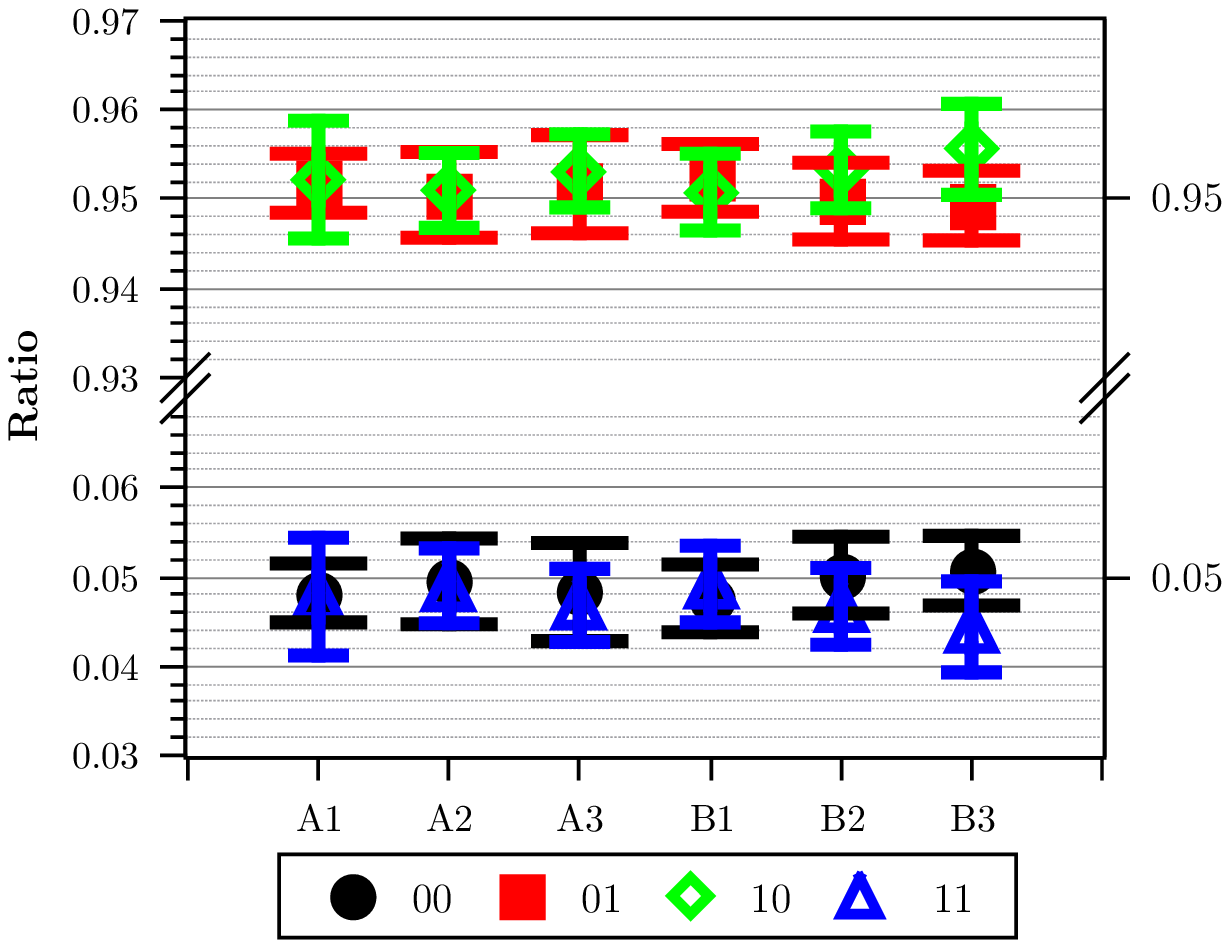}}}
    \subfigure[MMI\#4]{\resizebox*{0.35\textwidth}{!}{\includegraphics*{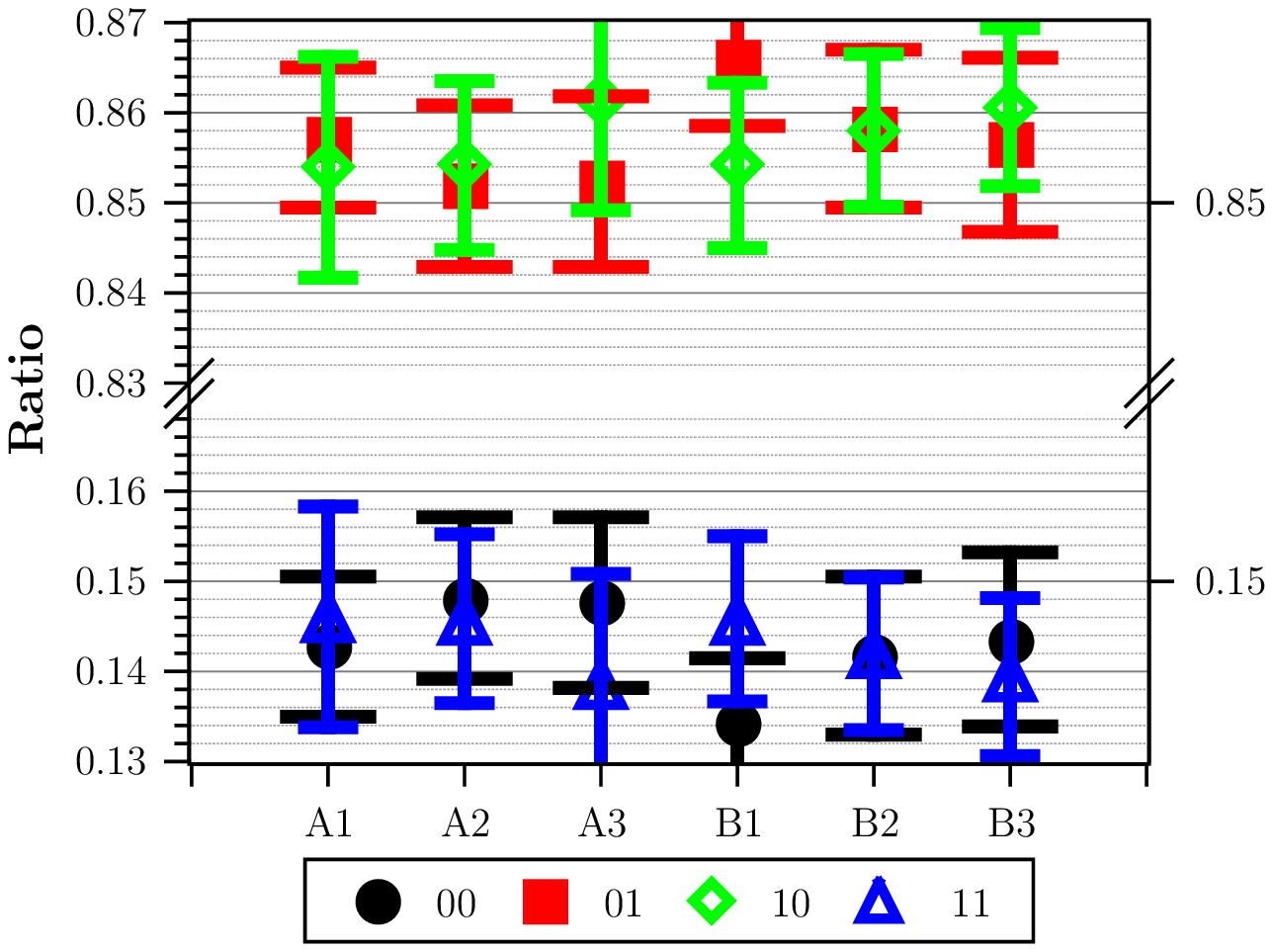}}}
    \caption{MMIs \#1 to \#4, average and standard deviation in $\lambda \in \left[1525,1575\right]$~nm for the coupling ratios, for wafers A y B and dies 1, 2, 3 within each wafer.}
    \label{fig:mmi_wafer_die_1}
}
\end{figure}

\begin{figure}
{\par\centering


    \subfigure[MMI\#5]{\resizebox*{0.35\textwidth}{!}{\includegraphics*{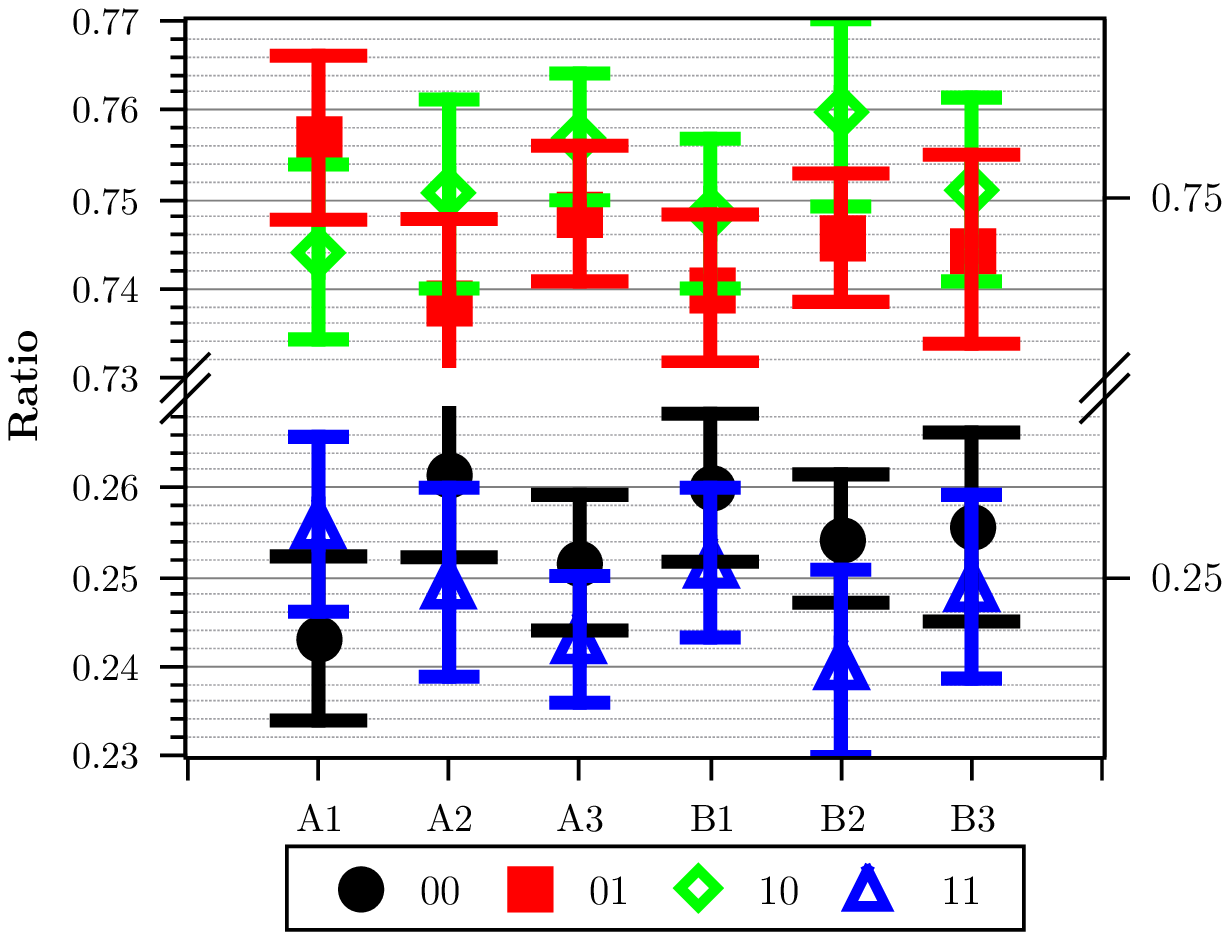}}}
    \subfigure[MMI\#6]{\resizebox*{0.35\textwidth}{!}{\includegraphics*{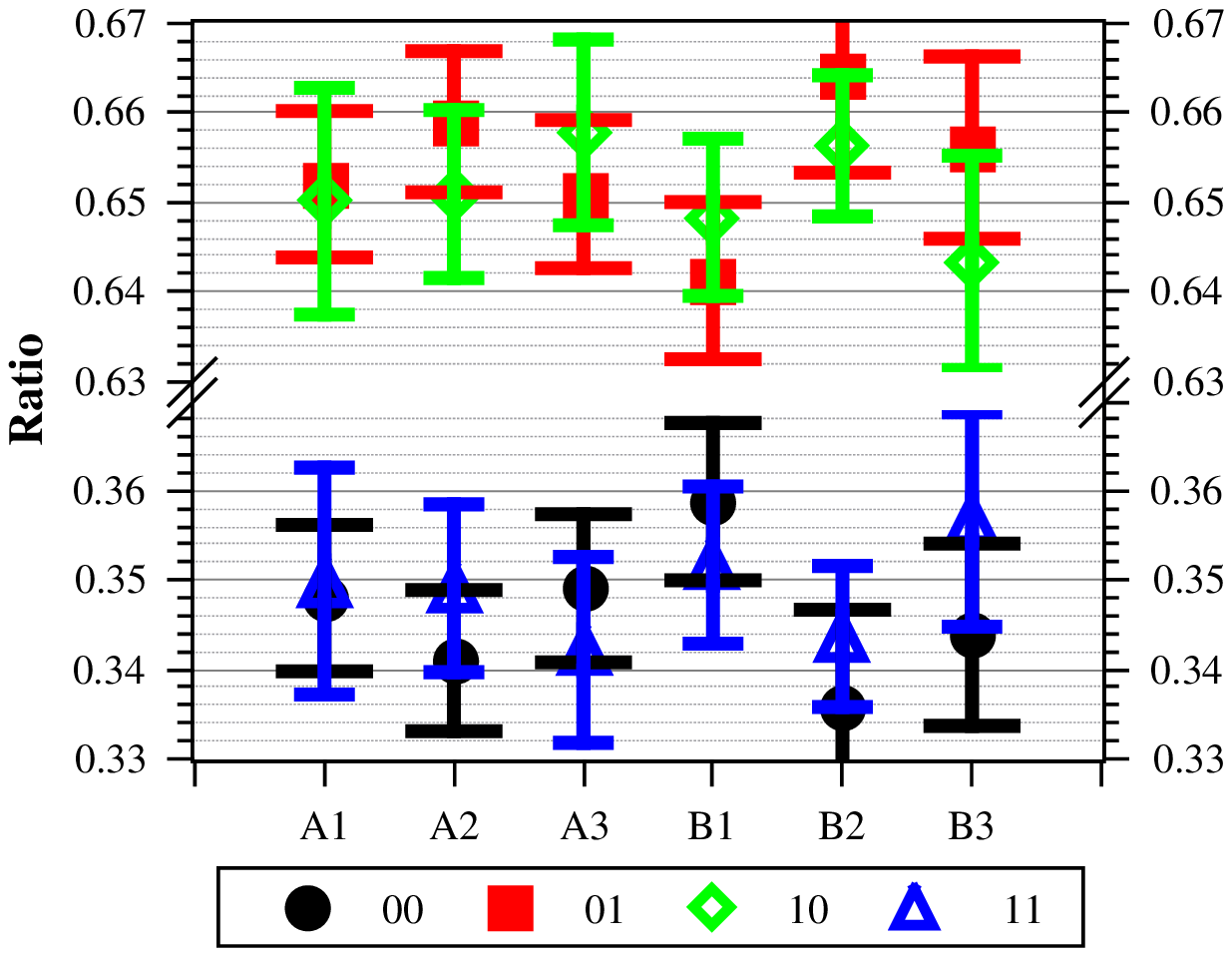}}}

    \subfigure[MMI\#7]{\resizebox*{0.35\textwidth}{!}{\includegraphics*{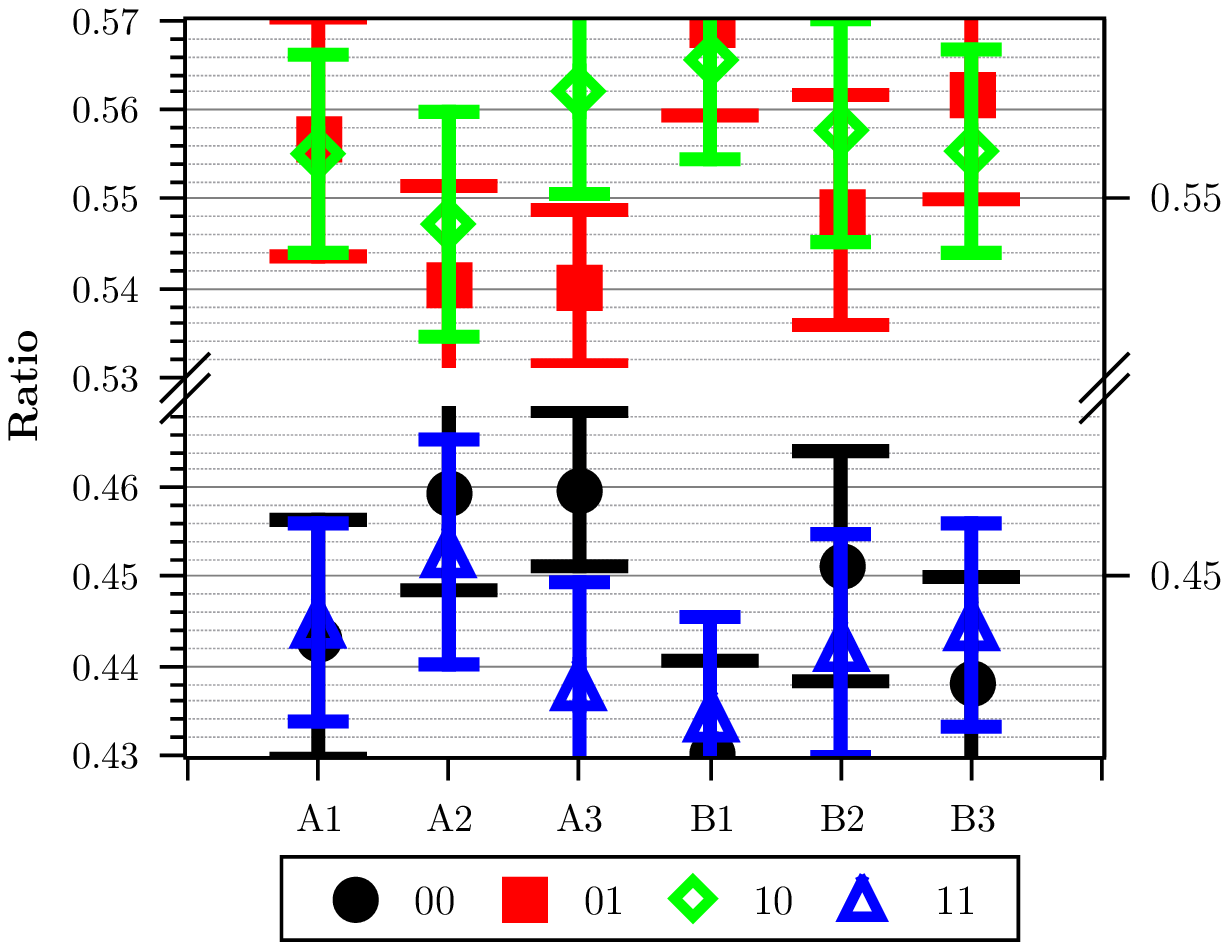}}}
    \subfigure[MMI\#8]{\resizebox*{0.35\textwidth}{!}{\includegraphics*{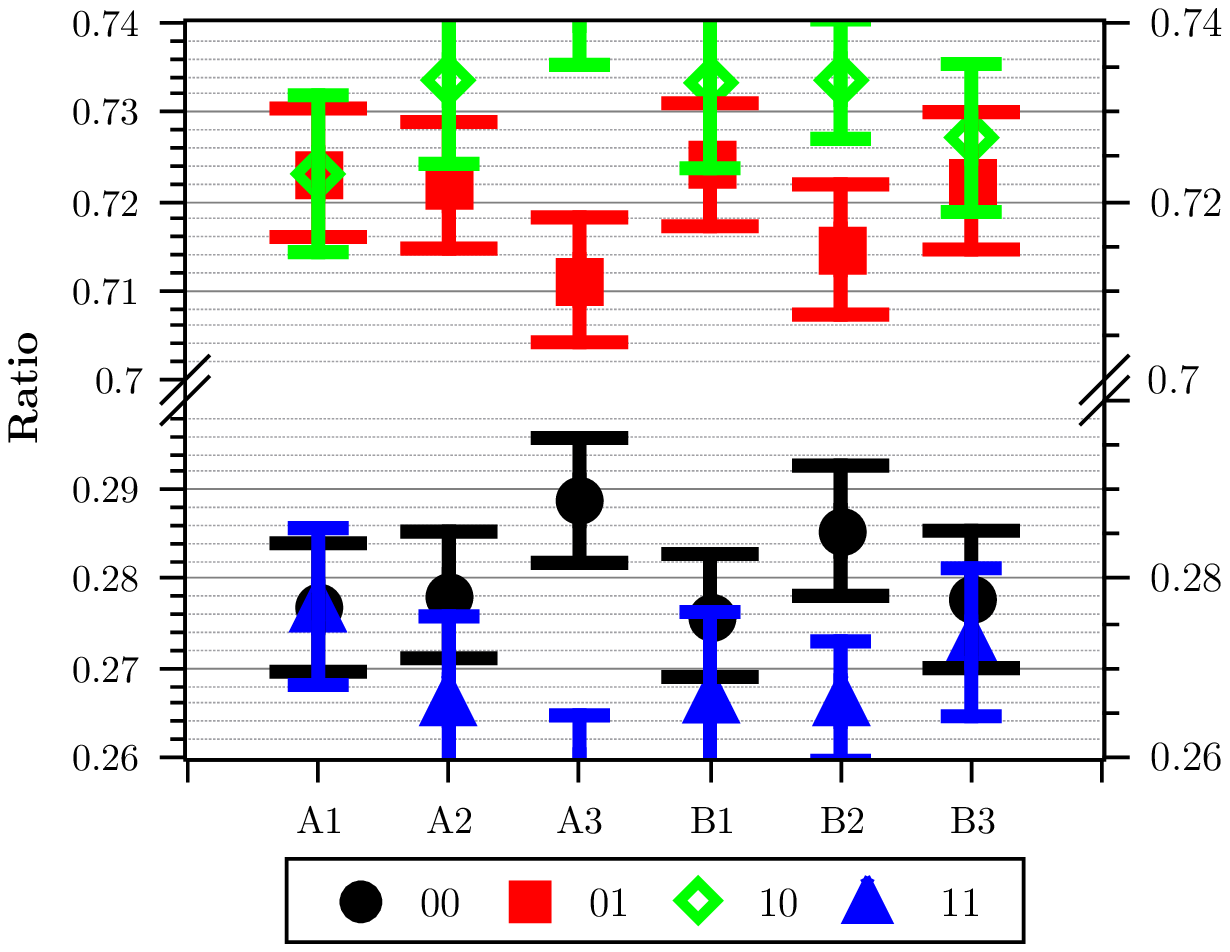}}}
    \caption{MMIs \#5 to \#8, average and standard deviation in $\lambda \in \left[1525,1575\right]$~nm for the coupling ratios, for wafers A y B and dies 1, 2, 3 within each wafer.}
    \label{fig:mmi_wafer_die_2}
}
\end{figure}

The coupling ratios for the MMIs were derived from the spectral traces measured at outputs 'm', from input 'l', termed $S_{l,m}(\lambda)$, as:
\begin{equation}
\label{eq:K}
K_{l,m}(\lambda) = \frac{S_{l,m}(\lambda)}{S_{l,0}(\lambda)+S_{l,1}(\lambda)}
\end{equation}
where l=0,1 and and m=0,1 label the input and output waveguides respectively, and with spectral traces in linear units. $K_{l,m}(\lambda)$ traces for MMI\#2 and MMI\#3 on die A1 are plotted in Fig.~\ref{fig:mmi_spectra}. The results show good agreement with target coupling ratios, where deviations are approximately in the range of $\pm$0.01. For the rest of the dies, similar spectral traces and deviations were obtained from both wafers, albeit the etch depth difference between wafer A and B. Note the additional etch depth of 30~nm in wafer B did not change significantly the results, which is in good agreement with the sensitivity analysis reported in \cite{halir}. The results for all the couplers per die and wafer are compiled in summary graphs given in Fig.~\ref{fig:mmi_wafer_die_1} and  Fig.~\ref{fig:mmi_wafer_die_2}, where the average coupling ratios and standard deviations in the wavelength range of the measurements are shown. 

MMI \#1 50:50 samples exhibited coupling constants around 0.5 with deviations in the whole wavelength range of less than $\pm$0.01 for all dies, except A1 and B3, where some imbalance is clearly appreciated. For MMIs \#2 to \#4 the graphs are given with broken axes, but with the same interval around the target coupling ratio. Comparing MMI \#2 and \#4, which have as target 85:15 splitting ratio, the performance of the first proved to be best for all dies. One might be tempted to attribute this to the fact that device \#2 is a canonical (rectangular body) design, whereas \#4 follows the widened body geometry shown in Fig.~\ref{fig:mmi_sketch}-(b), i.e. Type B Symmetrized. However MMI \#3 shown in Fig.~\ref{fig:mmi_wafer_die_1}-(c) exhibits very good performance, which might be misleadingly interpreted as MMI \#4 being a sub-optimal design. Therefore additional insight is provided in the following. Note the spectral traces $S_{l,m}(\lambda)$ are recorded at the two different outputs of the MMI, each one equipped with a FGC. Ideally both FGCs should have very similar performance. If this is not the case, a minor difference in the average power delivered from the FGCs changes Eq. (\ref{eq:K}) into:
\begin{equation}
\label{eq:Ks}
K_{l,m}^{\pm}(\lambda) = \frac{(1 \pm \frac{\Delta}{2} ) S_{l,m}(\lambda)}{(1 \mp \frac{\Delta}{2} )S_{l,0}(\lambda)+(1 \pm \frac{\Delta}{2} ) S_{l,1}(\lambda)}
\end{equation}
where $\Delta$ represents difference between average power delivered by each FGC. 
\begin{figure}
  {\par\centering
  \resizebox*{0.45\textwidth}{!}{\includegraphics*{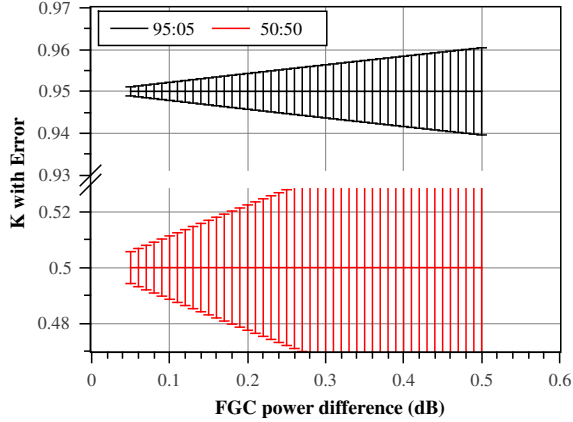}}
  \caption{Coupling ratio variation with average power difference delivered by the output FGCs, for nominal 50:50 and 95:05 coupling ratios.}
  \label{fig:Ks}
  }
\end{figure}

Hence, resorting to Fig.~\ref{fig:Ks}, the calculated coupling ratios, for a given difference in the performance of the FGCs, are more sensitive in the case of coupling ratios closer to 0.5. On the contrary, sensitivity to this issue is lower for more asymmetric couplers. This is clearly noticeable for MMI\#2 B1 in Fig.~\ref{fig:mmi_wafer_die_1}-(b), MMI\#7 A3 in Fig.~\ref{fig:mmi_wafer_die_2}-(c), MMI\#8 A3 in Fig.~\ref{fig:mmi_wafer_die_2}-(d) as well. 

Therefore the efficiency of the FGCs that may vary from one to other, not only between different dies, but inside the same die too \cite{hochberg_fgc}, is the most likely cause of the cases out of the general trends. Similar conclusions can be inferred for MMIs \#5 to \#8 from Fig.~\ref{fig:mmi_wafer_die_2}.

\subsection{Excess loss}
An estimation for the excess loss, EL, is derived combining the MMI measured spectra, $S_{l,m}(\lambda)$, with the spectra of reference straight waveguides, $S_{sw}(\lambda)$ as.
\begin{equation}
EL_l(\lambda) = 10\log_{10}\left[\frac{S_{sw}(\lambda)}{S_{l,0}(\lambda)+S_{l,1}(\lambda)}\right]
\end{equation} 
with all the magnitudes in linear units. The spectra of four straight waveguides per die, spanning the same length than the MMIs with input/outputs, were measured. The maximum value at each wavelength was calculated to obtain a single trace $S_{sw}(\lambda)$ in each die for normalization. The average excess loss for all wavelength was then calculated. The result needed to be corrected by adding 0.4~dB, meaning the deviations due to fiber alignments and differences in FGC performance are at least 0.4 dB. Consequently, the EL values obtained are best case. It is likely the actual value to be 0.2-0.4~dB larger (i.e. despite considering the FGCs having equal performance, the fibers measuring the straight waveguides might be slightly misaligned). Under these conditions, the numerical values for the excess loss average for all wafers, dies and MMIs are calculated. Consequently, one cannot derive and absolutely accurate value for the EL from these measurements. Otherwise, resorting to full in depth statistical analysis of a larger number of samples would be required, and it is out of the scope of this paper (see \cite{phd:dumon}, \cite{hochberg_fgc} and \cite{bogaerts_challenges} for reproducibility issues).
\begin{figure}
  {\par\centering
   \subfigure[]{\resizebox*{0.48\textwidth}{!}{\includegraphics*{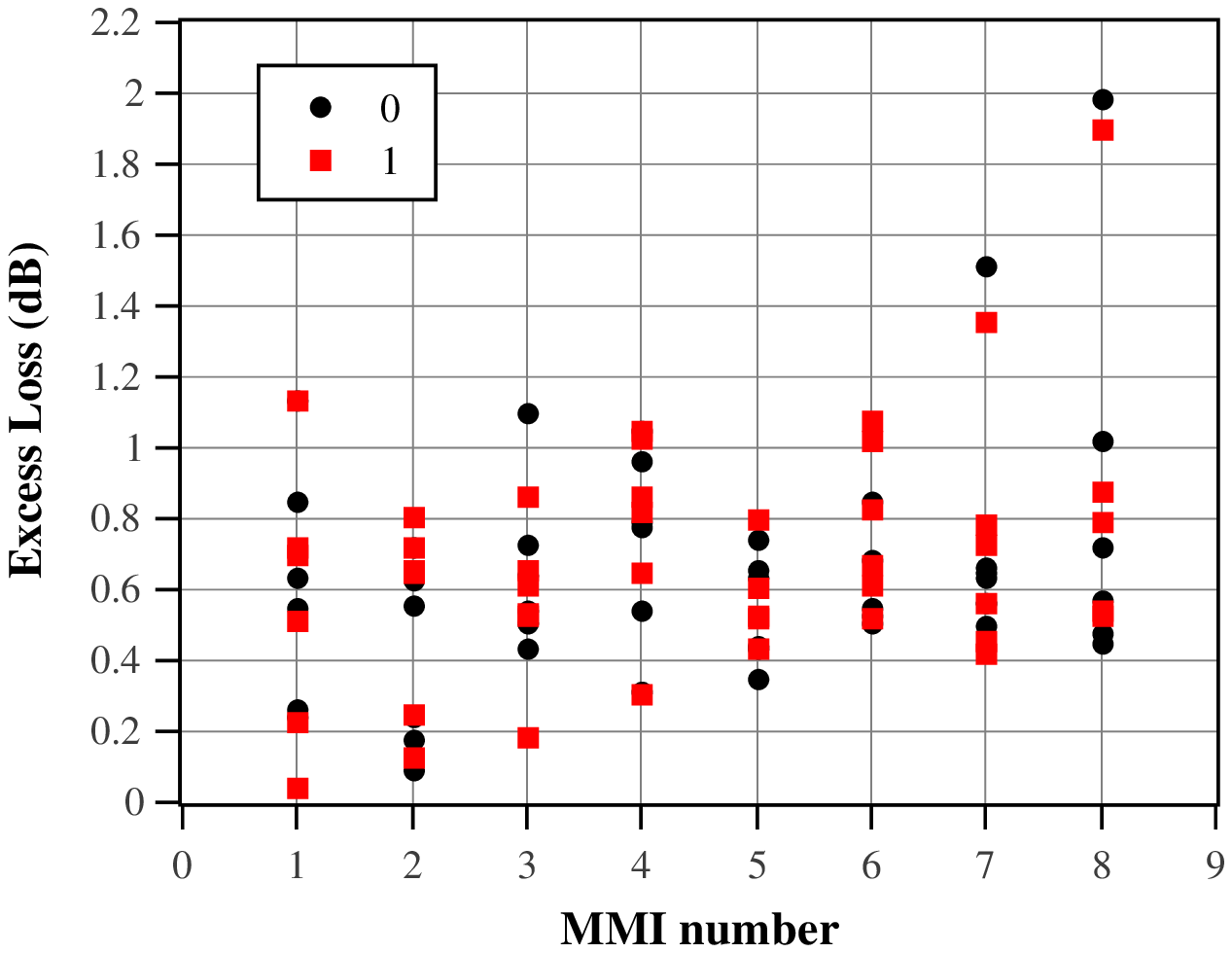}}}
   \subfigure[]{\resizebox*{0.48\textwidth}{!}{\includegraphics*{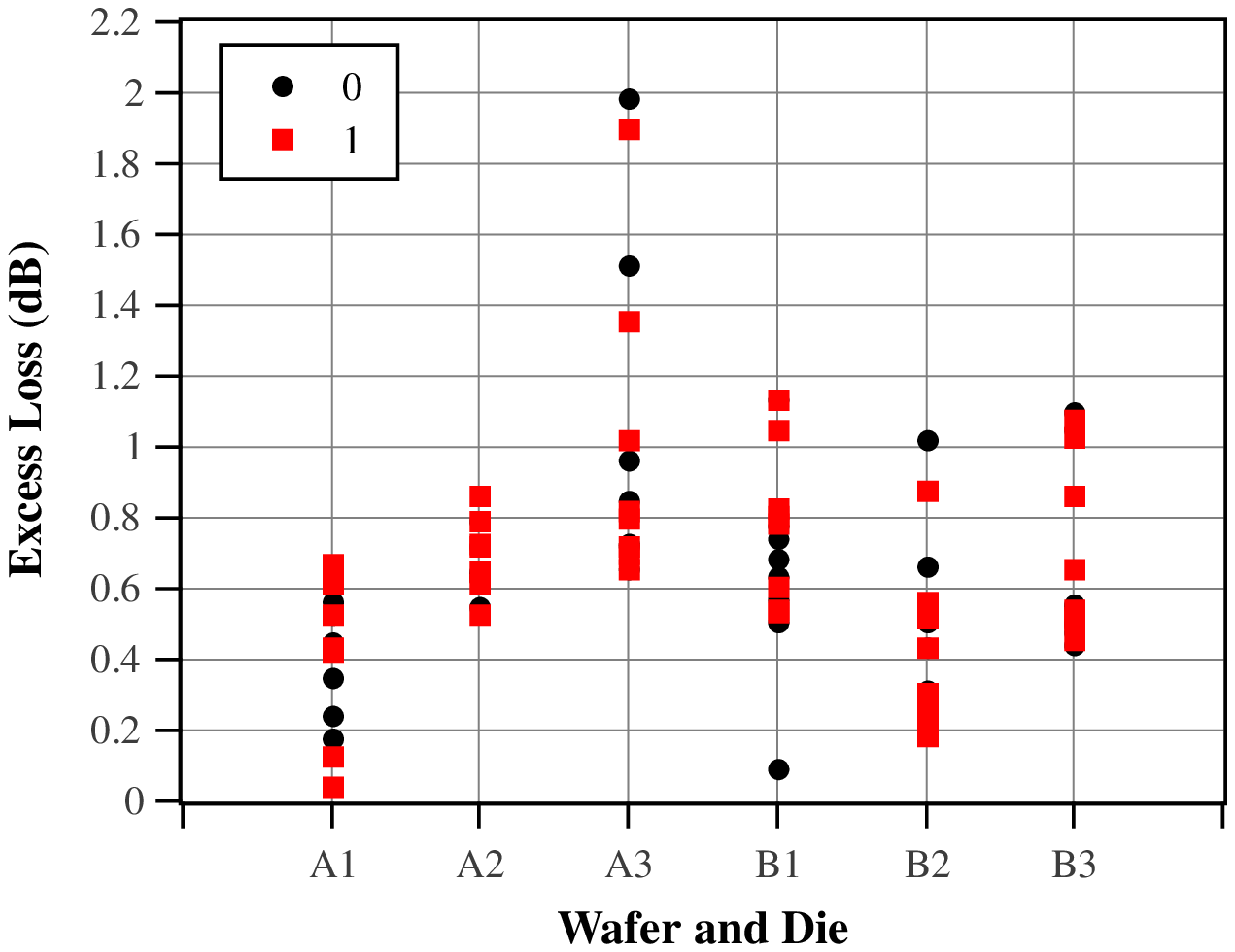}}}
  \caption{Stacked values for the estimated excess loss for (a) each MMI, all dies and (b) each die, all MMIs.}
  \label{fig:EL}
  }
\end{figure}
From these calculations, the interval on which the EL lies can be estimated at the sight of Fig.~\ref{fig:EL}. Panel (a) in the figure shows the estimated EL per MMI, i.e. all the dies for the each MMI. Panel (b) shows the estimated EL for all the MMIs in a die.  \ul{Except MMI\#8 at die~A3,} Fig.~\ref{fig:EL}-(a)\ul{, which had excess loss close to 2~dB, all the other devices/dies had losses in average around 0.6 dB, since die~A3 exhibited comparatively higher excess loss for all the MMIs, as can be inferred from} Fig.~\ref{fig:EL}-(b).


\section{Conclusion and outlook}
In this paper the design, fabrication and measurement of MMIs with arbitrary coupling ratio in Silicon-on-Insulator technology has been reported. The design methodology consisted on a combination of theoretical first guess and numerical optimization, using the Beam Propagation Method. The devices were fabricated in two different wafers, where the waveguides had different etch depths. Very good match between the design and experimental results was obtained in terms of the coupling ratio for the devices. All the coupling ratios were attained within the design wavelength range of 1525-1575~nm with deviations as low as $\pm$0.02. Minor deviations were attributed to the difference in the performance of the focusing grating couplers. \ul{Except for one die, the estimated average excess loss for the MMIs is around 0.6~dB. The statistical and reproducibility information on this paper can be readily incorporated by others to device, circuit and network on-chip simulation and design tools, in order to asses on more complex photonic chip circuits based on these MMIs.}
\section*{Acknowledgment}
The authors acknowledge financial support by the Spanish CDTI NEOTEC start-up programme, the Spanish MICINN project TEC2010-21337, acronym ATOMIC, the Spanish MINECO project TEC2013-42332-P, acronym PIC4ESP, project FEDER UPVOV 10-3E-492 and project FEDER UPVOV 08-3E-008. B. Gargallo acknowledges financial support through FPI grant BES-2011-046100.  J.S. Fandi\~no acknowledge financial support through FPU grant AP2010-1595.

\appendix*{Refractive indices}
The following wavelength dependence for the refractive indices of the materials was included in the solver given by Sellmeier equation\cite{weber}:

\begin{equation}
n^2\left(\lambda\right)=  1+\sum_{i=0}^N \frac{B_i\lambda^2}{\lambda^2-C_i}
\end{equation}
For Si and $\lambda \in \left[1.36-11\right]\mu$m the coefficients are B$_i$=\{10.66842933, 0.003043475, 1.54133408\} and C$_i$=\{0.3015116485$^2$, 1.13475115$^2$, 1104.0$^2$\}$\mu$m$^2$. For SiO$_2$ and $\lambda \in \left[0.21-3.71\right]\mu$m the coefficients are  B$_i$=\{0.6961663, 0.4079426, 0.1162414\} and C$_i$=\{0.0684043$^2$, 0.1162414$^2$, 9.896161$^2$\}$\mu$m$^2$.

\bibliographystyle{IEEEtran}
\bibliography{mmi}

\end{document}